 \documentclass[12pt,a4paper]{article}

\usepackage{graphicx}
\usepackage{graphics}
\usepackage{amssymb}
\usepackage{amsmath}
\usepackage{epsfig}
\usepackage{latexsym}
\usepackage{color}
\usepackage{rotating}
\usepackage{subfigure}

\newcommand{\bra}[2] {\mbox{}_{#2}\langle #1 |}
\newcommand{\ket}[2] {| #1 \rangle_{#2}}

\begin{document}

\title{\bf Entangling a nanomechanical resonator with a microwave field.}

\author{A. K. RINGSMUTH and G. J. MILBURN,\\
Department of Physics, School of Physical Sciences,\\
 The University
of Queensland, St Lucia, QLD 4072, Australia}
\maketitle
\begin{abstract}
We show how the coherent oscillations of a nanomechanical resonator can be entangled with a microwave cavity in the form of a superconducting coplanar resonator. Dissipation is included and realistic values for experimental parameters are estimated.
\vskip 2 truecm
{\em keywords}: nanomechanical, entanglement, superconducting.
\end{abstract}

\newpage

\section{Introduction}
Nano-mechanical resonators can now be fabricated with frequencies approaching  a gigahertz\cite{PT}. In this regime it is possible to consider using microwave transducers for their motion.  Superconducting coplanar microwave cavities \cite{coplanar} offer a fabrication geometry compatible with that used for nano-mechanical resonators and have recently enabled  a new class of experiments in circuit quantum electrodynamics in the strong coupling regime\cite{circuit-QED}.  In this paper we show how coherent states of a nano-electromechanical system (NEMS) may become entangled with the coherent states of a microwave field.  The scheme may be used to improve the sensitivity for weak force measurement using nano-mechanical resonators.

Oscillator coherent states are the quantum states that exhibit dynamics that is closet to the classical description of an oscillator. The correspondence extends to the case of two quadratically coupled oscillators initially prepared in coherent states as in that case the total system state, at all times factorises, into coherent states for each.  Entangled coherent states however are highly non classical and can only arise when the Hamiltonian is higher order than quadratic in the canonical variables\cite{Sanders92}. Such states can be used to get an quantum enhancement in the signal-to-noise ratio for weak force detection\cite{Munro} or in optical quantum computing\cite{Ralph}. There are proposals for generating these optically\cite{Song, Ritsch}.  Armour et al.\cite{Armour} have suggested a scheme for putting a single nano-mechanical resonator into a superposition of coherent states. 

\begin{figure}[th!]
\centering{\includegraphics[scale=0.7]{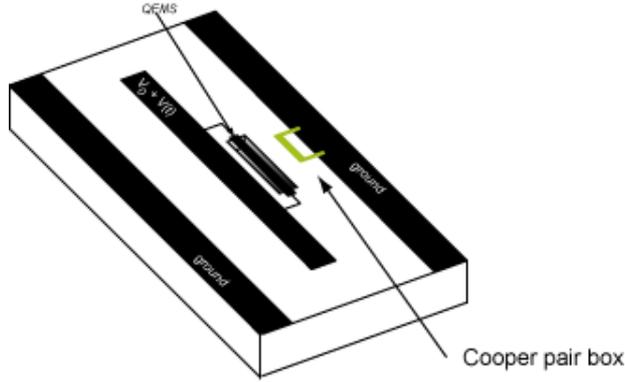}}
\caption{A nano-mechanical resonator is coupled to a microwave cavity via a Cooper pair box electric dipole. The nano-mechanical resonator carries a metal wire, is doubly clamped and suspended above a cut-away in the substrate. As it oscillates the DC bias voltage on a metal gate induces an AC bias voltage on the CPB.} 
\label{device}
\end{figure}

There are a number of ways that a nano-mechanical system may be coupled to a microwave cavity field. A possible experimental implementation is shown in figure \ref{device}. 
   The scheme we use here is based on that of Sun et al.\cite{Sun} in which the interaction is mediated by an off-resonance interaction with a Cooper pair box (CPB) qubit. It is similar to the scheme presented by Zhang et al. \cite{Zhang}.
 
The coupling between a Cooper pair box charge system and the microwave field of circuit QED is given by\cite{Blais}
\begin{equation}
H=4E_c\sum_{N}(N-n_g(t))^2|N\rangle\langle N|-\frac{E_J}{2}\sum_{N=0}\left (|N\rangle\langle N+1|+|N+1\rangle\langle N|\right )
\label{full_hamiltonian}
\end{equation}
where 
\begin{eqnarray*}
E_C & = & \frac{e^2}{2C_{\Sigma}}\\
n_g(t) & = & \frac{C_gV_g(t)}{2e}
\end{eqnarray*} 
with $C_\Sigma$ the capacitance between the island and the rest of the circuit, $C_g$ is the capacitance between the CPB island and the bias gate for the island, and $V_g(t)$ is the total voltage applied to the island by the bias gate composed of a DC field, $V_g^{(0)}$ and microwave field in the cavity, $\hat{v}(t)$. Thus we can write $V_g(t)=V_g^{(0)}+\hat{v}(t)$, where the hat indicates a quantisation of the cavity field. The Hamiltonian in Eq.(\ref{full_hamiltonian}) is written in the Cooper pair number basis. In this basis the electrostatic energy of the first term is quite clear. The Josephson energy term describes tunneling of single Cooper pairs across the junction. This term is more traditionally (i.e. in mean-field theory) written in the phase representation as $E_J\cos\theta$. The connection between these two representations is discussed in \cite{Spiller}.

We now allow for the possibility that the CPB (or its bias gate) is mounted on a nano-mechanical oscillator. This leads to a periodic modulation of the gate capacitance of the form 
\begin{equation}
C_g(t)=C_g(1-\hat{x}(t)/d)
\end{equation}
where $\hat{x}$ is the displacement operator for the nano-mechanical oscillator and $d$ is a typical length scale. Including both the time dependent cavity field and the modulated gate capacitance we can write
\begin{equation}
n_g(t)=n_g^{(0)}+\delta \hat{n}_g(t)
\end{equation}
where
\begin{equation}
\delta \hat{n}_g(t)=\frac{C_g}{2e}\hat{v}(t)-\frac{n_g^{(0)}}{d}\hat{x}(t)-\frac{C_g}{2ed}\hat{x}(t)\hat{v}(t)
\end{equation}

Using the usual restriction of the CPB Hilbert space to $N=0,1$, we can write the Hamiltonian as
\begin{equation}
H=-2E_C(1-2n_g^{(0)})\bar{\sigma}_z-\frac{E_J}{2}\bar{\sigma}_x-4E_C\delta\hat{n}_g(t)(1-2n_g^{(0)}-\bar{\sigma}_z)
\end{equation}
where $\bar{\sigma}_z=|0\rangle\langle 0-|1\rangle\langle 1|,\ \ \ \bar{\sigma}_x=|1\rangle\langle 0|+|0\rangle\langle 1|$. 
Define the bare CPB Hamiltonian as
\begin{equation}
H_{CPB}=-2E_C(1-2n_g^{(0)})\bar{\sigma}_z-\frac{E_J}{2}\bar{\sigma}_x
\end{equation}
and diagonalise it as 
\begin{equation}
H_{CPB}=\frac{\epsilon}{2}\sigma_z
\end{equation}
where 
\begin{equation}
\epsilon=\sqrt{E_J^2+[4E_C(1-2N_g^{(0)})]^2}
\end{equation}
and now the Hamiltonian takes the form,
\begin{equation}
H=\hbar \omega_r a^\dagger a +\frac{\epsilon}{2}\sigma_z-4E_C\delta\hat{n}_g(t)[1-2n_g^{(0)}-\cos\theta \sigma_z+\sin\theta\sigma_x]
\end{equation}
where we have now included the free Hamiltonian for the microwave cavity field, and
\begin{equation}
\theta=\arctan\left (\frac{E_J}{4E_C(1-2n_g^{(0)})}\right )
\end{equation}
Operating at the charge degeneracy point, $n_g^{(0)}=1/2$ so that $\theta=\pi/2$, the Hamiltonian becomes
\begin{eqnarray}
H & = &\hbar\omega_r a^\dagger a+\hbar\nu b^\dagger b+ \frac{\epsilon}{2}\sigma_z-4E_C\left (\frac{C_g}{2e}\hat{v}(t)-\frac{1}{2d}\hat{x}(t)-\frac{C_g}{2ed}\hat{x}(t)\hat{v}(t)\right )\sigma_x\\
 & = &  \hbar\omega_r a^\dagger a+\hbar\nu b^\dagger b+ \frac{\epsilon}{2}\sigma_z-\hbar g(a+a^\dagger)\sigma_x+\hbar \lambda(b+b^\dagger)\sigma_x+\hbar \chi(a+a^\dagger)(b+b^\dagger)\sigma_x
\end{eqnarray}
where we have now included the free Hamiltonian for the NEMS ($\hbar\nu b^\dagger b$), and
\begin{eqnarray}
\hbar g & = & e\frac{C_g}{C_\Sigma}\sqrt{\frac{\hbar\omega_r}{Lc}}\\
\hbar \lambda & = & 2E_C\frac{ x_{rms}}{d}\\
\hbar \chi & = & g\frac{x_{rms}}{d}
\end{eqnarray}
where $x_{rms}=\sqrt{\hbar/(2m\nu)}$ is the rms position fluctuations in the oscillator ground state. 

We now assume that the circuit resonance is at $\omega_r\approx 1$ GHz, and that the circuit and the  NEMS can be resonant. However, we detune the qubit from this resonant frequency by a few MHz \cite{Walraff}.  We will make the rotating wave approximation and take the Hamiltonian in the interaction picture to be 
\begin{equation}
H_I=\hbar \delta_a a^\dagger a+\hbar\delta_b b^\dagger b+\hbar g (a\sigma_++a^\dagger \sigma_-)+\hbar\lambda (b\sigma_++b^\dagger \sigma_-)
\end{equation}
with $\delta_a=\omega_r-\omega_0$ and $\delta_b=\nu-\omega_0$ are the detuning between the cavity resonance and the CPB and the NEMS and the CPB respectively, and $\hbar\omega_0=\epsilon$. 

We now assume that $\delta_a=\delta_b=\delta$ (that is, $\omega_r=\nu$) and that the time scale of interest is such that $t>>\delta^{-1}$. The effective interaction Hamiltonian is then 
\begin{eqnarray}
H_{I_{eff}}\equiv H = \left[\hbar\chi ~a^{\dag}a~ + \hbar\Omega~b^{\dag}b +\hbar \kappa\left(ab^{\dag}+a^{\dag}b\right)~\right]\sigma_z \label{Hab}
\end{eqnarray}
with $\chi = \frac{g^2}{\delta}$, $\Omega = \frac{\lambda^2}{\delta}$ and $\kappa = \frac{g\lambda}{\delta}$. If the CPB is in a superposition of its two states, we can use this term to entangle coherent states of the NEMS with the cavity field as we now show.

\section{\label{sec:non-dissipative}Non-dissipative case}
We first consider the case in which the NEMS, CPB and cavity are all non-dissipative. It will be shown that, in this case, unitary evolution under \eqref{Hab} results in entanglement between the vibrational modes of the NEMS and the cavity field. We will assume that the NEMS and cavity field  are in coherent states, $\ket{\alpha_0}{a}\ket{\beta_0}{b}$, at $t=0$. In the case that the CPB is in an eigenstate of $\sigma_z$, it can be shown that they remain in coherent states at all times under \eqref{Hab} and no entanglement can arise. However, if the CPB is prepared in an eigenstate of $\sigma_x$, entanglement is indeed possible.  We first define $\sigma_z|\pm\rangle=\pm|\pm\rangle$ and assume that the CPB is initially in its ground state, $|-\rangle$, to write
\begin{eqnarray}
\ket{\Psi(0)}{} &= \ket{\alpha_0}{a}\ket{\beta_0}{b}\ket{-}{},
\end{eqnarray}
Enacting a $\pi/2$ pulse on the CPB puts it into the $+$ eigenstate of $\sigma_x$, 
\begin{eqnarray}
\ket{\Psi(0)}{} &=\frac{1}{\sqrt{2}}~\ket{\alpha_0}{a}\ket{\beta_0}{b}\left(\ket{+}{}+\ket{-}{}\right).\label{sys2}
\end{eqnarray}

The state at later times may be written
\begin{eqnarray}
\ket{\Psi(t)}{} &= \frac{1}{\sqrt{2}}~\left(\ket{\Psi_+(t)}{ab}\ket{+}{} + \ket{\Psi_-(t)}{ab}\ket{-}{}\right). \label{evol}
\end{eqnarray} 
where 
\begin{equation}
|\Psi_\pm(t)\rangle_{ab}=e^{\mp itH_{\pm}}|\alpha_0\rangle_a\otimes|\beta_0\rangle_b
\end{equation}
with $H=\chi a^\dagger a+\Omega b^\dagger b+\kappa(ab^\dagger +a^\dagger b)$. Clearly the dynamics for $\sigma_z=-1$ is the time reversed dynamics for $\sigma_z=+1$.  If however the CPB is in a superposition of the two eigenstates, we have the fascinating result of a superposition of the dynamics and the time reversed dynamics.  

Using the well known fact that the Hamiltonians $H_\pm$ transform coherent states into coherent states,
the overall system state at time $t$ can be written as, 
\begin{eqnarray}
\ket{\Psi(t)}{} &= \frac{1}{\sqrt{2}}\left (\ket{\alpha(t)}{a}\ket{\beta(t)}{b}|+\rangle+\ket{\alpha(-t)}{a}\ket{\beta(-t)}{b}|-\rangle\right )
\label{sys}
\end{eqnarray} 
where
\begin{eqnarray}
\alpha(t) & = & e^{-i\omega t}\left [\left (\frac{\kappa\beta_0}{R}-\frac{\alpha_0(\Delta-R)}{2R}\right ) e^{-iRt/2}+\left (\frac{\alpha_0(\Delta+R)}{2R}-\frac{\kappa\beta_0}{R}\right )e^{iRt/2}\right ]\\
\beta(t) & = & e^{-i\omega t}\left [\left (\frac{\kappa\alpha_0}{R}+\frac{\beta_0(\Delta+R)}{2R}\right ) e^{-iRt/2}-\left (\frac{\beta_0(\Delta-R)}{2R}+\frac{\kappa\alpha_0}{R}\right )e^{iRt/2}\right ]
\label{total-state}
\end{eqnarray}
and $\omega=(\Omega+\chi)/2,\ \ \Delta=\Omega-\chi,\ \ \ \ R=\sqrt{\Delta^2+4\kappa^2}$. 
Note that for either case, $\sigma_z=\pm1$, we have the conservation law $|\alpha(t)|^2+|\beta(t)|^2=|\alpha_0|^2+|\beta_0|^2$. 
The vibrational modes of the NEMS and the cavity field have become entangled. To detect the coherence implicit in this entanglement we need to readout $\sigma_x$ on the CPB.

The usual way to do this is to again subject the CPB to a $\pi/2$ rotation pulse, followed by a readout of $\sigma_z$ for the CPB. This is equivalent to a direct readout of $\sigma_x$ on the state in Eq. (\ref{total-state}).   After the $\pi/2$ pulse, the system is then in the state,
\begin{eqnarray}
|\psi'(t)\rangle & = & -\frac{1}{2}\left (\ket{\alpha(t)}{a}\ket{\beta(t)}{b}-\ket{\alpha(-t)}{a}\ket{\beta(-t)}{b}\right )|+\rangle\\
& & +\frac{1}{2}\left (\ket{\alpha(t)}{a}\ket{\beta(t)}{b}+\ket{\alpha(-t)}{a}\ket{\beta(-t)}{b}\right )|-\rangle.
\end{eqnarray}
It is then clear that a readout of the CPB state will project the cavity and NEMS into entangled coherent states, or `cat states'.  This is very similar to the NIST ion trap experiment to produce a cat state for the vibrational motion of the ion\cite{NIST}. The coherence implicit in the cat appears in the probability for the CPB measurement results. For example, the probability to find the CPB in the state $|-\rangle$ is
\begin{equation}
P_-(t)=\frac{1}{2}\left [1+\Re\left (\langle \alpha(t)|\alpha(-t)\rangle\langle \beta(t)|\beta(-t)\rangle\right )\right ],
\end{equation}
which may be written as
\begin{equation}
P_-(t)=\frac{1}{2}\left \{1+\exp\left [-\frac{1}{2}\left (|\alpha(t)-\alpha(-t)|^2+|\beta(t)-\beta(-t)|^2\right )\right ]\cos[\Phi(t]\right \}
\end{equation}
where the interference fringes as a function of time are determined by the phase
\begin{equation}
\Phi(t)=\Im\{\alpha^*(t)\alpha(-t)+\beta^*(t)\beta(-t)\}
\end{equation}
while the visibility of the interference is suppressed exponentially in the square of the distance between the coherent amplitudes and the time-reversed coherent amplitudes.  This has a simple geometric interpretation given in figure \ref{fig-3}. 

\begin{figure}[th!]
\includegraphics[scale=1.0]{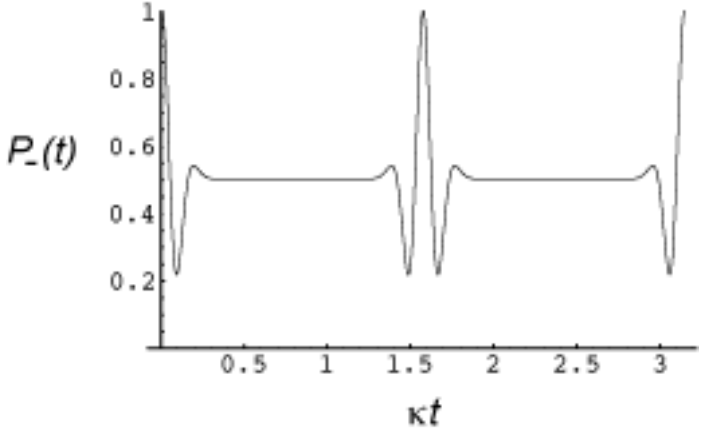}
\caption{ A plot of the probability to find the CPB in the initial state $|-\rangle$ as a function of $t$, with $B=4.0, \kappa=1.0$.} 
\label{fig-2}
\end{figure}

\begin{figure}[th!]
\includegraphics[scale=0.7]{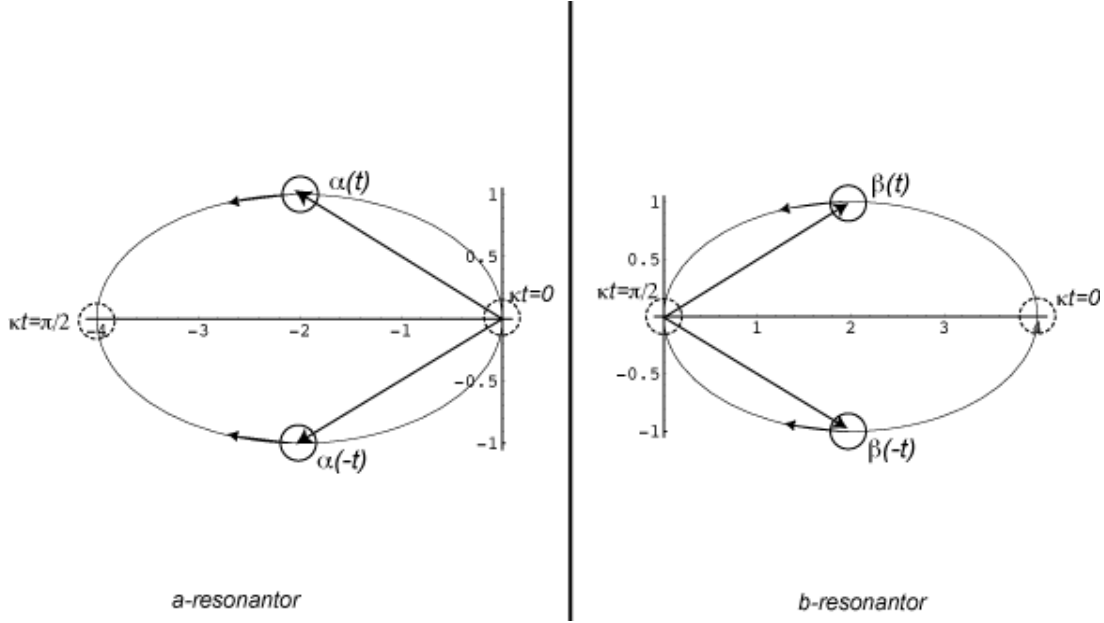}
\caption{A graphical phase space representation of the generation of entanglement between the resonators and the CPB, and its manifestation as interference in $P_-(t)$. On the left(right) we represent the forward and reversed dynamics of the complex amplitude of $a(b)$-resonator. The parameters are the same as in figure \ref{fig-2}. } 
\label{fig-3}
\end{figure}

As a specific example consider the symmetric case in which $\Omega=\chi=\kappa$ so that $\Delta=0$. We further suppose that the nano-mechanical resonator is prepared with coherent amplitude $\beta_0=B$, taken as real, while the cavity is cooled to the ground state $\alpha_0=0$. In this case we find that
\begin{eqnarray}
\alpha(t) & = & -iBe^{-i\kappa t}\sin\kappa t \\
\beta(t) & = & B e^{-i\kappa t}\cos\kappa t
\end{eqnarray}
so
\begin{equation}
P_-(t)=\frac{1}{2}\left \{1+\exp\left [-B^2\sin^2 2\kappa t \right ]\cos[B^2\sin(4\kappa t)/2]\right \}
\end{equation}
In figure \ref{fig-2} we plot this function versus time for the case $B=4.0, \kappa=1.0$. 

The collapse and revival of the interference can be interpreted geometrically, using a phase space representation, as shown in figure \ref{fig-3}.  Here we plot the orbits in the complex plane, parameterised by time, of the complex amplitudes $\alpha(\pm t)$ (on the left), and  $ \beta(\pm t)$ (on the right).  The solid and dotted circles represent the minimum uncertainty circles for the coherent state. The dotted circle indicates a time at which we have complete overlap of the amplitude with its time-reversed form. At this point we expect from previous studies to see maximum interference\cite{DFW-GJM}.  This agrees with the result in figure \ref{fig-2}.  However, when the two coherent amplitudes do not overlap, the interference is exponentially suppressed and $P_-(t)$ is very close to $0.5$.
This geometric interpretation in terms of interference in phase space is analogous to the picture of Knight and co workers\cite{Knight}  for explaining the collapse and revivals in the Jaynes-Cummings model.

\section{\label{dissipative}Dissipative case}
We now consider the case in which both the NEMS and cavity lose energy to their surroundings, which we treat as a zero temperature heat bath. This may be a reasonable assumption for the microwave cavity but is doubtful for a nano-electromechanical oscillator as they are currently operated, however it will  at least capture the general features of decoherence as a best-case situation.  We treat the damping in each case using the quantum optics master equation (weak coupling, rotating wave approximation). In the case of both oscillators approaching GHz resonant frequencies this approach is likely to be a reasonable approximation. In practice the decay rate for the nano-mechanical oscillator is likely to be greater than that for the microwave field.  For simplicity we will ignore the qubit decay. The qubit itself is subject to both dephasing and spontaneous emission, both of which occur at rates of the order of 1 MHz\cite{Schuster}. As we now show, the rate of decay of coherence between the superposed coherent states is proportional to the  modulus square of the initial coherent amplitudes and thus likley to be very much greater than decoherence due to qubit decay.  However, if necessary, the qubit decoherence can be included trivially as it simply causes an exponential decay, at fixed rate, of the off-diagonal term in the qubit basis. In general, the state of the system at any time is now described by a density matrix, 

\begin{eqnarray}
\rho(t) &= \frac{1}{2}\left (\rho_{++}(t)\ket{+}{}\bra{+}{} + \rho_{--}(t)\ket{-}{}\bra{-}{} + \rho_{+-}(t)\ket{+}{}\bra{-}{} + \rho_{-+}(t)\ket{-}{}\bra{+}{} \right )
\label{density}
\end{eqnarray}
and the dynamics is given by the master equation,
\begin{eqnarray}
\frac{d\rho}{dt} &= -i[H_I,\rho]+\gamma_a \mathcal{D}[a]\rho+\gamma_b \mathcal{D}[b]\rho 
\label{master}
\end{eqnarray}
where, for some operator $\mathcal{A}$, 
\begin{eqnarray}
\mathcal{D}[\mathcal{A}]\rho&= \mathcal{A}\rho \mathcal{A}^\dag - \frac{1}{2}(\mathcal{A}^\dag \mathcal{A}\rho+\rho \mathcal{A}^\dag \mathcal{A})
\end{eqnarray}
and $\gamma_a$ and $\gamma_b$ are respectively the dissipation rate constants for the cavity and the NEMS.

In the case that the qubit is in an eigenstate of $\sigma_z$ at $t=0$, the resulting irreversible dynamics  takes coherent states to coherent states, as does the reversible dynamics. This is a direct result of assuming zero temperature heat baths, and the state remains pure under irreversible evolution. We are thus led to adopt the ansatz that, as in the non-dissipative case, the cavity and NEMS will remain in coherent states, $|\alpha_{\pm}(t)\rangle_a|\beta_{\pm}(t)\rangle_b|\pm\rangle$ throughout unitary evolution. However, it is no longer true that the $\sigma_z=-1$ case is the time reversed version of $\sigma_z=+1$ case.  We can then write for the diagonal terms, $\rho_{++}(t)\ket{+}{}\bra{+}{},\  \rho_{--}(t)\ket{-}{}\bra{-}{}$,
\begin{eqnarray}
\rho_{++} (t) &= & |\alpha_+(t),\beta_+(t)\rangle\langle\alpha_+(t),\beta_+(t)|\\
\rho_{--} (t) &= & |\alpha_-(t),\beta_-(t)\rangle\langle\alpha_-(t),\beta_-(t)|
\end{eqnarray}
where we have defined $|\alpha,\beta\rangle=|\alpha\rangle_a\otimes|\beta\rangle_b$. 

Inserting this into \eqref{master} and equating like terms gives us a system of ordinary differential equations for the coherent amplitudes:
\begin{eqnarray}
\dot{\alpha}_{\pm} & = &  \mp i(\chi \alpha_{\pm}+\kappa \beta_{\pm})-\frac{ \gamma_a }{2}\alpha_{\pm}\\
\dot{\beta}{\pm} & = &  \mp i(\kappa \alpha_{\pm}+\Omega \beta_{\pm})-\frac{\gamma_b}{2}\beta_{\pm}  \\
\frac{d}{dt}(|\alpha_{\pm}|^2 + |\beta_{\pm}|^2) & = &-( \gamma_a |\alpha_{\pm}|^2 +\gamma_b |\beta_{\pm}|^2)  \label{DE5}
\end{eqnarray}
with initial conditions, 
\begin{eqnarray}
{\alpha_{\pm}}(0) &= \alpha_0 \label{alpha_0}\\
{\beta_{\pm}}(0) &= \beta_0. \label{beta_0}
\end{eqnarray}
Note that  $\alpha_-(t)=\alpha_-(t)^*,\ \ \beta_-(t)=\beta_+(t)^*$. 

We now turn to the off-diagonal term,  $\rho_{-+}(t)$.
Let us adopt the ansatz that $\rho_{-+}$ has the form
\begin{eqnarray}
\rho_{-+}(t) &= & f(t)\ket{\alpha_-(t),\beta_-(t)}{}\bra{\beta_+(t),\alpha_+(t)}{} \label{rho01}
\label{cornice}
\end{eqnarray}
with $f(t=0)=1$.  The function $f(t)$ controls the degree of coherence in the state due to the damping. For example, it is easy to see that the CPB coherence is given by
\begin{equation}
\langle \sigma_x\rangle(t)=\Re \left [f(t)\right ]
\end{equation}
In the  protocol discussed for the non-dissipative case,  it will determine $P_-(t)$ as
\begin{equation}
P_-(t)=\frac{1}{2}\left [1+\Re\left (f(t) \langle \alpha_+(t)|\alpha_-(t)\rangle\langle \beta_+(t)|\beta_-(t)\rangle\right )\right ]
\end{equation}

By substituting Eq. (\ref{cornice}) into the master equation and equating like terms we arrive at the same system of  differential equations as previously and in addition, 
\begin{eqnarray}
\frac{df}{dt}-\frac{1}{2} f \frac{d}{dt}(|\alpha_-|^2+|\beta_-|^2+|\alpha_+|^2+|\beta_+|^2)&= f\left(\gamma_a\alpha_-\alpha_+^*+\gamma_b\beta_-\beta_+^*\right) 
\label{DE52}
\end{eqnarray}
Using Eqs. (38-42) this may be written as 
\begin{eqnarray}
\frac{df}{dt} &= f~\mathcal{G}(t)  \label{consistent}
\end{eqnarray}
where 
\begin{equation}
\mathcal{G}(t)=  \gamma_a\left (\alpha_-\alpha_+^*-|\alpha_-|^2\right )+\gamma_b\left (\beta_-\beta_+^*-|\beta_-|^2\right).\label{G}
\end{equation}
Using Eqs. (38-40) we can write
\begin{equation}
\mathcal{G}(t)=\gamma_a(\alpha_-(t))^2+\gamma_b(\beta_-(t))^2+\frac{d}{dt}\left (|\alpha_-(t)|^2+|\beta_-(t)|^2\right )
\end{equation}
So,
\begin{equation}
 f(t) = \exp\left(\left (|\alpha_-(t)|^2+|\beta_-(t)|^2\right )-\left (|\alpha_0|^2+|\beta_0|^2\right )+\int_0^t dt' \gamma_a(\alpha_-(t'))^2+\gamma_b(\beta_-(t'))^2\right).
 \label{Bob}
\end{equation}
As the equations of motion for $\alpha_{\pm},\beta_{\pm}$ are a simple linear system, this may be easily computed by direct integration. The solutions are
\begin{eqnarray}
\alpha_-(t) & = & U e^{-(\gamma_+-i(\omega-w/2))t}+Ve^{-(\gamma_+-i(\omega+w/2))t}\\
\beta_-(t) & = & X e^{-(\gamma_+-i(\omega-w/2))t}+Ye^{-(\gamma_+-i(\omega+w/2))t}
\end{eqnarray}
where
\begin{eqnarray}
U & = & \frac{\alpha_0}{w}\left [(w+\Delta)/2+i(\gamma_b-\gamma_a)/4\right ]-\kappa\frac{\beta_0}{w}\\
V & = & \frac{\alpha_0}{w}\left [(w-\Delta)/2-i(\gamma_b-\gamma_a)/4\right ]+\kappa\frac{\beta_0}{w}\\
X & = & \frac{\beta_0}{w}\left [(w-\Delta)/2-i(\gamma_b-\gamma_a)/4\right ]-\kappa\frac{\alpha_0}{w}\\
Y & = & \frac{\beta_0}{w}\left [(w+\Delta)/2+i(\gamma_b-\gamma_a)/4\right ]+\kappa\frac{\alpha_0}{w}
\end{eqnarray}
with $w=\sqrt{4\kappa^2-(\gamma_a-\gamma_b+2i\Delta)^2/4}$ and $\gamma_+=(\gamma_a+\gamma_b)/4$. 
We then find that
\begin{eqnarray}
{\cal G}(t) & = & \gamma_a\left [x_+(t)U^2+x_-(t)V^2+2UVy(t)\right ]\\
				& &	+\gamma_b\left [x_+(t)X^2+x_-(t)Y^2+2XYy(t)\right ]\\ 
				& &	+(|\alpha_-(t)|^2-|\alpha_0|^2)+(|\beta_-(t)|^2-|\beta_0|^2)
\end{eqnarray}
where
\begin{eqnarray}
x_\pm & = & \frac{1-e^{-2t(\gamma_+-i(\omega\mp w/2)}}{2(\gamma_+-i(\omega\mp w/2))}\\
y(t) & = & \frac{1-e^{-2t(\gamma_+-i\omega)}}{2(\gamma_+-i\omega)}.
\end{eqnarray}					

\begin{figure}[ht]
	\centering
		\includegraphics[width=0.5\textwidth]{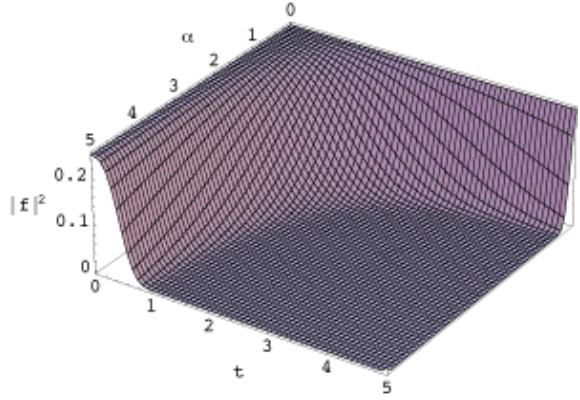}
	\caption{Decoherence as seen in the decay of $|f(t)|^2$ using the short time approximation, with $\alpha=\beta, \gamma t<<1,\gamma=0.1, \chi=\Omega=0, \kappa = 1$.}
	\label{fig:simplef}
\end{figure}

\begin{figure}[ht]
	\centering
		\includegraphics[width=0.5\textwidth]{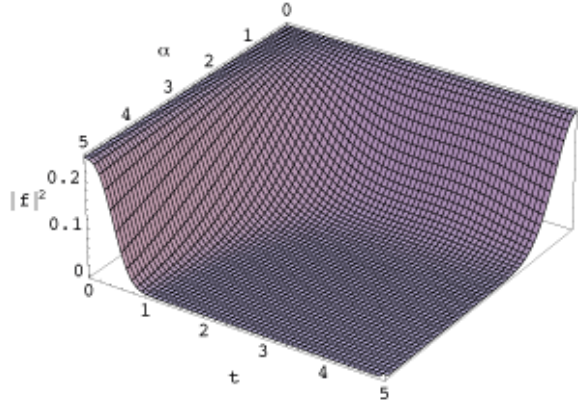}
	\caption{Decoherence as seen in the decay of $|f(t)|^2$ using the full solution, with  $\alpha=\beta, \gamma t<<1,\gamma=0.1, \chi=\Omega=0, \kappa = 1$.}
	\label{fig:generalf}
\end{figure}

To gain an understanding of how decoherence works at short times,  we consider the  simple symmetric case: $\chi=\Omega=\kappa$, $\gamma_a = \gamma_b = \gamma$ with the initial conditions, $\alpha_0=0,\ \ \beta_0=B$. 
In this case we find that
\begin{eqnarray}
\alpha_-(t) & = & ie^{-(\gamma-i\kappa)t}B\sin\kappa t\\
\beta_-(t) & = & e^{-(\gamma-i\kappa)t}B\cos\kappa t.
\end{eqnarray}
Thus
\begin{equation}
f(t)=\exp\left [-\frac{B^2}{2}\left (1-e^{-\gamma t}\right)+\frac{\gamma B^2}{2}\left (\frac{1-e^{-\gamma t+4i\kappa t}}{\gamma-4i\kappa}\right )\right ]
\end{equation}
For short times ($\gamma t<<1$), this can be approximated by
\begin{equation}
f(t)= \exp\left [-\frac{4B^2}{3}\gamma\kappa^2 t^3+iB^2\kappa \gamma t^2\right ]
\end{equation}
Thus the visibility of the interference decays at short times via the prefactor
\begin{equation}
|f(t)|^2=\exp\left [-\frac{8B^2}{3}\gamma\kappa^2 t^3\right ]
\end{equation}
a plot of which is given in fig. \ref{fig:simplef}. 
The cubic dependence on $t$ has been seen in other models of this kind\cite{DFW-GJM}. It is not unexpected. Coherence cannot be corrupted until it is created. Under unitary evolution, short time dynamics is always quadratic to lowest order in $t$, which then decays at a rate proportional to $t$ for short times. The dependence of the decay rate on $B^2$ is also expected, as previous studies have shown that the coherence decays at a rate proportional to the square of the separation of the amplitudes of the superposed coherent states\cite{DFW-GJM}.  Fig.\ref{fig:generalf} shows the general numerical solution for this case, where $\alpha_0$ is again assumed to be real.
Comparing figures (\ref{fig:simplef}, \ref{fig:generalf}) we see that for these parameters the short time approximation is quite good. 

\section{Discussion and conclusion.}
We now make some optimistic estimates of the parameters that might be achieved in an experiment.  From Walraff et al.\cite{Walraff} we find that  $g\approx 6$MHz and $ E_C/\hbar \approx 5$GHz. We will take the detuning between the two resonators and the CPB to be of the order of one MHz. We further suppose $\nu=1$ GHz  with a mass of  $m=10^{-21}$kg \cite{Roukes} and thus $x_{rms}\approx 10^{-2}$nm. With advanced fabrication it should be possible to achieve $d\approx 20$nm and thus $\lambda=2(E_c/\hbar)(x_{rms}/d)\approx 3MHz$.  It thus seems reasonable to expect that  $\lambda/g=0.5$.  We scale time so that $\chi=1$, so that in dimensionless units, $\kappa/\chi=0.5,\ \Omega/\chi=0.25$. The quality factor for the cavity resonance can be quite high, perhaps as high as $10^4$. With a resonance frequency of  1GHz, this gives in dimensionless units that $\gamma=10^{-3}$. The quality factor for the mechanical resonance is not as good, so we take it to be a factor of 10 smaller so that $\gamma_b=10^{-2}$. This might be achievable for a very small nano-mechanical resonator as considered here. Let us take the microwave resonator and mechanical resonator to have low coherent amplitudes $\alpha_0=\beta_0=2.0$, which should be achievable at milli Kelvin temperatures.
\begin{figure}[ht]
	\centering
		\includegraphics[width=0.5\textwidth]{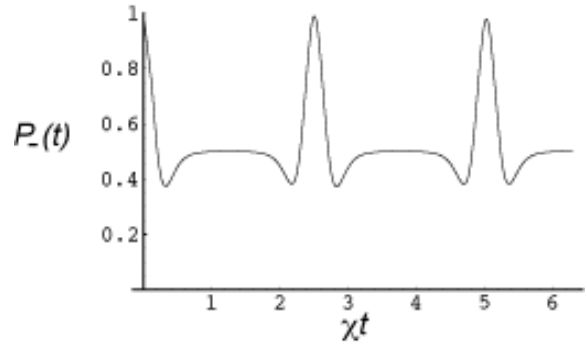}
	\caption{The probability to find the CPB in the ground state using the protocol in section \ref{sec:non-dissipative} including dissipation, with  $\alpha=\beta=2, \ \ \gamma_a=0.001,\ \gamma_b=0.01, \chi=1.0,\ \  \Omega=0.25, \kappa = 0.5$.}
	\label{fig-4}
\end{figure}
In figure (\ref{fig-4}) we plot the probability to find the CPB in the ground state using the protocol of section \ref{sec:non-dissipative}. Interference effects are clearly evident giving some hope that the quantum entanglement implicit in this result could be observed. 

We have presented a method to entangle the coherent motion of a nano-mechanical resonator with a coherent state of a microwave cavity field. In so far as the nano-mechanical resonator is a mesoscopic mechanical system, this is already a fascinating prospect similar in many ways to the two-slit interference of a fullerne molecule\cite{Zeilinger}. However, it may also be of some use. Entangled coherent states can be used to improve the sensitivity for weak force measurements\cite{Munro}. As weak force detection is one of the primary motivations for building nano-mechanical systems, this may prove a useful way to improve performance. 

\section*{acknowledgements.}
We thank Andrew Doherty for advice. This work was supported by the Australian Research Council.

\newpage

\section*{Figure captions}
\begin{enumerate}
\item A nano-mechanical resonator is coupled to a microwave cavity via a Cooper pair box electric dipole. The nano-mechanical resonator carries a metal wire, is doubly clamped and suspended above a cut-away in the substrate. As it oscillates the DC bias voltage on a metal gate induces an AC bias voltage on the CPB.\\
\item A plot of the probability to find the CPB in the initial state $|-\rangle$ as a function of $t$, with $B=4.0, \kappa=1.0$.  \\
\item A graphical phase space representation of the generation of entanglement between the resonators and the CPB, and its manifestation as interference in $P_-(t)$. On the left(right) we represent the forward and reversed dynamics of the complex amplitude of $a(b)$-resonator. The parameters are the same as in figure \ref{fig-2}. \\
\item Decoherence as seen in the decay of $|f(t)|^2$ using the short time approximation, with $\alpha=\beta, \gamma t<<1,\gamma=0.1, \chi=\Omega=0, \kappa = 1$.\\
\item Decoherence as seen in the decay of $|f(t)|^2$ using the full solution, with  $\alpha=\beta, \gamma t<<1,\gamma=0.1, \chi=\Omega=0, \kappa = 1$.\\
\item The probability to find the CPB in the ground state using the protocol in section \ref{sec:non-dissipative} including dissipation, with  $\alpha=\beta=2, \ \ \gamma_a=0.001,\ \gamma_b=0.01, \chi=1.0,\ \  \Omega=0.25, \kappa = 0.5$.
\end{enumerate}
\end{document}